\documentclass[useAMS,usenatbib]{mn2e}
\usepackage{graphicx}

\title[Discovery of Raman-scattered lines in LHA 115-S 18]{Discovery of Raman-scattered lines in the massive luminous emission-line star LHA 115-S 18\thanks{Based on observations taken with 1) ESO Telescopes at La Silla Observatory, Chile, under the programmes 076.D-0609(A) and 080.A-9200(A), and 2) du Pont Telescope at Las Campanas Observatory, Chile, under the programme CNTAC 2008-02, PI: Barb\'a}}
\author[A. F. Torres, M. Kraus, L. S. Cidale, R. Barb\'a, M. Borges Fernandes and E. Brandi]{A. F. Torres$^{1, 2}$\thanks{E-mail:
andy@carina.fcaglp.unlp.edu.ar (AFT)}, M. Kraus$^{3}$, L. S. Cidale$^{1, 2}$, R. Barb\'a$^{4}$, M. Borges Fernandes$^{5}$ 
\newauthor 
and E. Brandi$^{1, 6}$\\
$^{1}$Departamento de Espectroscop\'ia, Facultad de Ciencias Astron\'omicas y Geof\'isicas, Universidad Nacional de La Plata, Paseo del Bos-\\
que S/N, La Plata, B1900FWA, Buenos Aires, Argentina \\
$^{2}$Instituto de Astrof\'isica de La Plata (CCT La Plata - CONICET, UNLP), Paseo del Bosque S/N, La Plata, B1900FWA, Buenos Ai-\\
res, Argentina\\ 
$^3$ Astronomick\'y \'ustav, Akademie v\v{e}d \v{C}esk\'e republiky, Fri\v{c}ova 298, 251 65 Ond\v{r}ejov, Czech Republic \\
$^4$ Instituto de Ciencias Astron\'omicas, de la Tierra y del Espacio (ICATE - CONICET), Casilla 467, 5400 San Juan, Argentina \\
$^5$ Observat\'orio Nacional, Rua General Jos\'e Cristino 77, 20921-400 S\~ao Cristov\~ao, Rio de Janeiro, Brazil\\
$^6$ Comisi\'on de Investigaciones Cient\'ificas de la Provincia de Buenos Aires (CIC), Calle 526 entre 10 y 11, Buenos Aires, Argentina}
\begin{document}

\date{Accepted 2012 September 4. Received 2012 September 4; in original form 2012 June 2.}

\pagerange{\pageref{firstpage}--\pageref{lastpage}} \pubyear{}

\maketitle

\label{firstpage}

\begin{abstract}
LHA 115-S 18 is a very peculiar emission-line star exhibiting the B[e] phenomenon. Located in the Small Magellanic Cloud,
 its spectrum shows features of an extremely wide range of excitation and
 ionization stages, extending from highly ionized atomic lines (Si\,{\sc iv}, C\,{\sc iv}, He\,{\sc ii}) in 
the UV and optical regions to molecular emission bands of CO and TiO in the 
optical and IR regions. The most distinguishing spectral characteristic of LHA 115-S 18 is the high variability detected in the He\,{\sc ii} $\lambda$4686 emission line, which can be a very conspicuous or completely invisible feature. 

 In this work, we report on another peculiarity of LHA 115-S 18. From
high-resolution optical spectra taken between 2000 and 2008, we discovered the
appearance and strengthening of two emission features at $\lambda$6825 \AA\, and $\lambda$7082 \AA\,, which we identified as Raman-scattered lines. This is the first time these lines have been detected in the spectrum of a massive luminous B[e] star. As the classification of LHA 115-S 18 is highly controversial, we discuss how the discovery of the appearance of Raman-scattered lines in this peculiar star might help us to solve this puzzle.
\end{abstract}


\section{Introduction}

Understanding the B[e] phenomenon is a constant challenge in stellar astrophysics. This phenomenon characterizes B-type stars exhibiting in the optical spectral range a rich low-excitation emission-line spectrum dominated by Balmer lines, narrow permitted emission lines of predominantly singly ionized metals, forbidden emission lines of [Fe\,{\sc ii}] and [O\,{\sc i}], and a strong near or mid-infrared excess that provides evidence for hot circumstellar dust. The classification for B[e] stars proposed by \citet{lam98} reveals that the B[e] phenomenon can be observed in emission-line stars of different evolutionary stages. It does not only exist in post-main sequence stars, as B[e] supergiant and compact planetary nebulae B[e] stars, but also in pre-main sequence objects as HAeB[e] stars. Symbiotic systems, which consist of a hot star, typically a white dwarf, and a cool red giant companion, also show the B[e] phenomenon at some stage in their evolution. Furthermore, a group of B[e] objects remains unclassified, exhibiting properties that are common to more than one class.

LHA 115-S 18 is a really striking object displaying the B[e] phenomenon. Located at the Small Magellanic Cloud, this extreme emission-line star was identified as S 18 by \citet{hen56} and classified as a B[e] supergiant by \citet{zic89}, who derived the following stellar parameters: $T_{\mathrm {eff}}$ = 25\,000 K, $\log\,g$ = 3.0, $E(B-V)$ = 0.4, $L_*$ = 3.0 - 4.6 $\times$ 10$^5$ $L_{\odot}$, $R_*$ = 33 - 36 $R_{\odot}$, and a ZAMS mass of $M\sim$ 35 - 40 $M_{\odot}$. Although there is a general agreement that LHA 115-S 18 is a massive post-main sequence object, its nature remains unclear, it has also been proposed as possible luminous blue variable (LBV), $\alpha$ Cygni variable, planetary nebulae or symbiotic star \citep{lin55, san78, mor96, mas01, van02}. Its spectrum is characterized by a great variety of features, as well as an extreme variability. Features of an extremely wide range of excitation and ionization stages, extending from highly ionized atomic lines (Si\,{\sc iv}, C\,{\sc iv}, He\,{\sc ii}) in the UV and optical regions to molecular emission bands of TiO and CO in the optical and IR regions, have been reported \citep{azz79, sho82, sho87, zic89, mor96, not96}. A strong IR-excess has also been detected and attributed to thermal reradiation by circumstellar dust (Kastner et al. \citeyear{kas10}, Bonanos et al. \citeyear{bon10}). In addition, strong spectral variations have been observed in the H\,{\sc i} Balmer series lines, which have shown pure emission-line as well as P Cygni-type profiles \citep{lin55, azz75, azz81, zic89}. However, the most puzzling variability has been reported for He\,{\sc ii} $\lambda$4686 line, which can be a very conspicuous feature at some epochs, and at others be completely absent \citep{san78}.  

In this Letter we report the discovery of Raman-scattered lines in the peculiar emission-line star LHA 115-S 18. This is the first time these lines have been detected in the spectrum of a massive luminous B[e] star. As the stellar classification of LHA 115-S 18 is highly controversial, we discuss how the discovery of the appearance of Raman-scattered lines might help us to solve this puzzle. The Letter is organized as follows. In Section \ref{sec:observations}, we describe our spectroscopic observations. In Section \ref{sec:results}, we present a brief description of the spectrum of LHA 115-S18 and its variability, and analyse the presence of the Raman-scattered lines. In Section \ref{sec:discussion} we discuss our results in the context of possible scenarios related to some different stellar classifications of LHA 115-S 18 and present our conclusions. A more detailed analysis of our data of LHA 115-S 18 will be presented in an upcoming publication (Torres et al., in preparation).

\section{Observations}\label{sec:observations}

Between 2000 and 2008 we obtained high-resolution optical spectra of LHA 115-S 18. The majority of our data set was taken with the Fiber-fed Extended Range Optical Spectrograph (FEROS), which was attached to the 1.52-m telescope until the end of 2002 and then it was moved to the 2.2-m telescope, both at the European Southern Observatory (ESO) in La Silla, Chile. FEROS is a bench-mounted echelle spectrograph, which provides data with a resolving power $R \sim$ 55\,000 (in the region around 6000 \AA) and a spectral coverage from 3600 to 9200 \AA. The observations were carried out on 2000 October 13, 2001 November 24, 2005 December 10, and 2007 October 3 and 4. The obtained signal-to-noise (S/N) ratio in the 5500 \AA\, region is between 10 and 25. The reduction process was performed using the FEROS standard on-line reduction pipeline. 

Two additional spectra were obtained on 2008 November 13, with the echelle spectrograph at the 2.5-m du Pont Telescope at Las Campanas Observatory (LCO) in Chile. The chosen instrumental configuration gave a spectral resolution of $R \sim$ 45\,000 and the wavelength range covered extends from 3600 to 9200 \AA. The S/N ratio in the 5500 \AA\, region is 65. The spectra were reduced using standard IRAF\footnote{IRAF is distributed by the National Optical Astronomy Observatories, which are operated by the Association of Universities for Research in Astronomy, Inc., under cooperative agreement with the National Science Foundation.} tasks. Bias subtraction, flat-field normalization, and wavelength calibration were performed.

\section{Results}\label{sec:results}

\begin{figure*}
 \includegraphics[angle=270,scale=0.57,bb= 80 80 500 700]{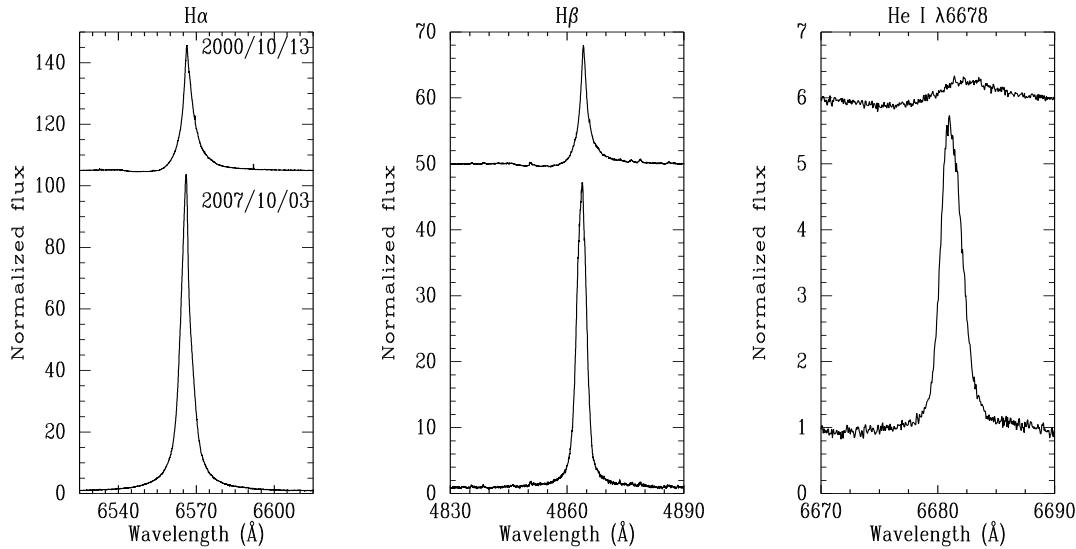}
  \caption{H$\alpha$, H$\beta$ and He\,{\sc i} $\lambda$6678 lines seen in the high-resolution spectra of LHA 115-S 18 obtained in 2000 (P Cygni line profile) and 2007 (pure emission line profile). The spectra have been normalized in flux and shifted for display purposes.}
  \label{balmer-lines-and-hei-2000-2007}
\end{figure*}

LHA 115-S 18 presents an emission-line spectrum completely dominated by hydrogen lines as well as permitted and forbidden transitions of neutral and ionized elements, mainly Fe\,{\sc ii}, [Fe\, {\sc ii}], Ti\,{\sc ii}, Cr\,{\sc ii}, [O\,{\sc i}], [O\,{\sc iii}], etc., as it was previously reported in the literature. Our complete data set of high-resolution spectra show no photospheric absorptions. A comparative detailed study of the spectra in the years 2000 and 2001 reveals significant changes with respect to those of 2005, 2007 and 2008. Line-profile variations in hydrogen, helium, iron and oxygen transitions are clearly detected. As an example, in Figure \ref{balmer-lines-and-hei-2000-2007}, we show the changes observed in the line profile of H$\alpha$, H$\beta$ and He\,{\sc i} $\lambda$6678 between 2000 (top plot) and 2007 (bottom plot). In the year 2000 the spectra are mainly characterized by the presence of P Cygni-type profiles in the {H\,{\sc i}} lines down to H$\zeta$, with differences in velocity between the emission peak and the center of the absorption feature of about 750 km s$^{-1}$. {He\,{\sc i}} lines also present P Cygni-type profiles, with a shallow absorption component. The {He\,{\sc ii}} emission line at $\lambda$4686 \AA\, is completely absent. The spectra obtained in 2001 are noisy, some lines loose the P Cygni-profile structure and the He\,{\sc ii} $\lambda$4686 line remains absent. From 2005, huge spectral changes become evident, mainly, the Balmer and {He\,{\sc i}} lines are strong and observed in pure emission with FWHMs of about 180, 30 and 80 km s$^{-1}$ for H$\alpha$, H$\beta$ and He\,{\sc i} $\lambda$6678 lines, respectively. The intensity of these lines increases from 2005 to 2007 followed up by a decrease in 2008. The most noticeable spectral changes detected in 2005 and subsequent years are the appearance of the He\, {\sc ii} $\lambda$4686 line, displaying a strong emission, and in the red part of the spectrum, the emergence of a broad emission bump which we identified with the Raman-scattered emission line at $\lambda$6825 \AA\,. This broad emission feature has not been previously reported in LHA 115-S 18. 

 \subsection{The detection of Raman-scattered lines}

\begin{figure*}
\includegraphics[angle=270,scale=0.53,bb= 80 60 470 720]{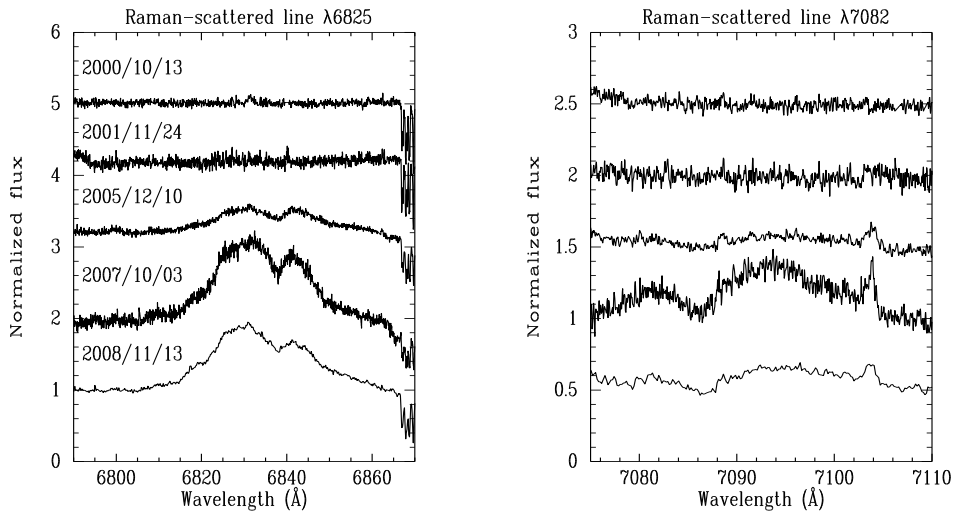}
  \caption{Temporal evolution of the Raman-scattered lines of LHA 115-S 18 between the years 2000 and 2008. The appearance of the emission feature at $\lambda$6825 \AA\, is observed in 2005 (left panel). The very broad but weak butterfly-like shape of the $\lambda$7082 Raman-scattered line becomes visible for the first time in 2007 (right panel). The spectra have been normalized in flux and shifted for display purposes. Dates of the spectra displayed in the right panel are the same as those of the left panel.}
  \label{s18-raman-lines}
\end{figure*}

In the spectra of LHA 115-S 18 obtained in 2007, a broad and intense feature around $\lambda$6830 \AA\, called our attention. We found that this structure is similar to the well-known emission bump at $\lambda$6825 \AA\, frequently detected in symbiotic systems \citep{all80,sch89,bir00a}. The presence of the $\lambda$6825 emission feature is often observed together with another similar emission structure at $\lambda$7082 \AA\,, which is weaker and sometimes undetectable. In symbiotic stars, these emission features are usually called Raman-scattered lines. Our data also reveal the presence of the second emission structure in the spectra taken in 2007 and 2008. Therefore, the confident detection of both features in the spectrum of LHA 115-S 18 enables us to identify them as possible Raman-scattered lines. In addition, it is worth mentioning the high variability displayed by these lines which are absent in the spectra obtained in the years 2000 and 2001 (see Figure \ref{s18-raman-lines}).   

Raman-scattered lines have been identified as due to Raman scattering of photons of the O\,{\sc vi} $\lambda$$\lambda$1032, 1038 resonance lines by neutral hydrogen \citep{sch89,bir00a}. Unlike Rayleigh scattering, Raman scattering is inelastic, that is the emitted photon has a different frequency to that of the incoming photon, and hydrogen is left on an altered quantum mechanical state. In this case, an incident O\,{\sc vi} photon excites hydrogen from its ground state 1$s\,{^{2}}\mbox{S}$ to an intermediate state near 3$p\,{ ^{2}}\mbox{P}$, from where the Raman scattered photon is emitted, leaving the hydrogen atom in the excited state 2$s\,{^{2}}\mbox{S}$. At present, Raman-scattered O\,{\sc vi} lines have only been observed in symbiotic stars, which have the appropriate physical conditions for efficient Raman scattering. In these objects, O\,{\sc vi} photons are most probably produced in the ionized region near the hot component and then inelastically scattered in a dense neutral hydrogen region near the red giant.

Unfortunately, there are no far-UV spectra of LHA 115-S 18 available in the literature to search for the O\,{\sc vi} $\lambda$$\lambda$1032, 1038 emission lines responsible for the optical Raman-scattered lines. So, to confirm a Raman-scattered origin of the emission bumps at $\lambda$$\lambda$6825, 7082 \AA\, we had to look for line transitions of highly ionized elements in the visible spectral region that could reveal the presence of a hot radiation source.  

According to \citet{lee04}, who reported a correlation between the equivalent widths of He\,{\sc ii} $\lambda$4686 line and those of $\lambda$6825 Raman-scattered line in the symbiotic system AG Dra, the intensity of He\,{\sc ii} line traces the variations in the number of O\,{\sc vi} photons since it forms close to the hot component. A similar dependence on the equivalent widths of H$\beta$ and the already mentioned Raman-scattered line was reported. 

\begin{figure*}
\includegraphics[angle=270,scale=0.52,bb= 75 55 510 650]{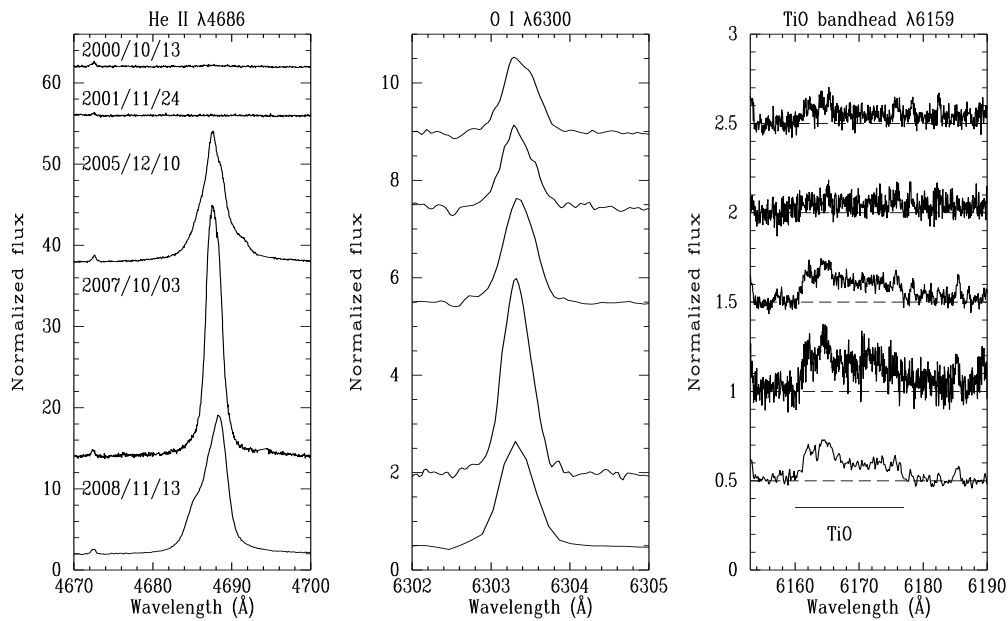}
  \caption{Temporal variation of He\,{\sc ii} $\lambda$4686 (left) and [O\,{\sc i}] $\lambda$6300 (centre) emission lines, and the wide emission feature at $\lambda$6159 \AA\, (right), probably related to a TiO molecular band. The spectra have been normalized in flux and shifted for display purposes. In the right panel, the continuum level is represented by a dashed line. Dates of the spectra displayed in the central and right panels are the same as those of the left panel.}
  \label{s18-heii-4686-and-tio-band-and-oi-6300}
\end{figure*}

In LHA 115-S 18, we also observed a strong direct relation between the intensities of the He\,{\sc ii} $\lambda$4686 and Raman-scattered lines. The appearance of both features occurred at the same epoch, strengthening and weakening their intensities in a similar way (see Figures \ref{s18-raman-lines} and \ref{s18-heii-4686-and-tio-band-and-oi-6300}). The same behaviour is displayed by the H$\beta$ line, except in the spectra of the years 2000 when Raman-scattered lines are absent and hydrogen lines exhibit P Cygni-type profiles. The variability detected in the lines accounts for a change in the physical conditions of the circumstellar environment: the increase in He\,{\sc ii} $\lambda$4686 line intensity reveals an enhancement of ionizing photons of a hot source, while the morphological change in H\,{\sc i} lines and the increase in intensity could indicate a mass ejection event that favoured a later mass density increment in the neutral hydrogen region. Moreover, if we considered that the emission of O\,{\sc i} is expected to arise from regions in which hydrogen is neutral, due to the about equal ionization potentials of H and O \citep{kra05, kra07}, then the behaviour of [O\,{\sc i}] lines would trace the changes in H\,{\sc i} lines. Figure \ref{s18-heii-4686-and-tio-band-and-oi-6300} shows an increase in the [O\,{\sc i}] $\lambda$ 6300 line emission towards the year 2007 and its subsequent decrease. This behaviour agrees with that observed in the Balmer lines.

\section{Discussion  and conclusions}\label{sec:discussion}

In this Letter we have reported for the first time the discovery of the appearance of Raman-scattered lines in the massive luminous B[e] star LHA 115-S 18.

The presence of Raman-scattered lines in LHA 115-S 18 is detected simultaneously with the appearance of He\,{\sc ii} $\lambda$4686, showing a close relationship between their intensities, as well as with those of H\,{\sc i} lines, indicating the presence of a hot ionizing radiation source and an enhancement of the neutral hydrogen circumstellar region, respectively. The observed spectral variability should be related to specific physical conditions and geometry of the circumstellar environment. 

Up to now, Raman-scattered O\,{\sc vi} lines have only been detected in symbiotic systems, which consist of a late-type giant star, a more compact star hot enough to produce ionizing photons, and a common circumbinary nebula. The original spectroscopic criterion outlined by \citet{ken86} to identify these objects is mainly based on the presence of a red continuum with absorption features, associated to the cool star, such as Ca\,{\sc i}, Ca\,{\sc ii}, TiO, VO, etc., and a blue continuum with strong emission lines of H\,{\sc i} and He\,{\sc i}, and in some cases dominated by high-excitation emission lines of He\,{\sc ii}, [Fe\,{\sc vii}], [Ne\,{\sc v}], etc. A more relaxed classification criterion was suggested by \citet{bel00} in which the presence of the $\lambda$6825 emission feature is enough to be a symbiotic star, even if no features of the cool star are found. However, in symbiotic stars the $\lambda$6825 Raman-scattered line is observed only in high excitation symbiotic systems showing [Ne\,{\sc v}] and [Fe\,{\sc vii}] lines \citep{sch89,bir00b}.

In LHA 115-S 18, the existence of Raman-scattered lines in its spectrum might indicate a symbiotic nature. However, [Fe\,{\sc vii}] emission lines are absent ([Ne\,{\sc v}] emission lines are out of the spectral range covered by our data), and there are no signs of red giant absorption features. Moreover, the presence of a wide emission feature at $\lambda$6159 \AA\,, typically seen in B[e] supergiants and probably related to a TiO molecular band \citep{zic89}, would point out to a different stellar nature (see Figure \ref{s18-heii-4686-and-tio-band-and-oi-6300}, right panel).

On the basis of our new observational evidence, a symbiotic B[e] scenario seems not to be the most plausible. Therefore, we briefly discuss two alternative scenarios: 1) a B[e] supergiant with a main-sequence hot star accreting mass, and 2) a LBV object. B[e] supergiants may harbour dense circumstellar discs of neutral hydrogen and dust, as in LHA 120-S 127 \citep{kas06, kra07}, where scattering Raman might occur if an intense source of O\,{\sc vi} photons exists. In this context, the extreme variability of \mbox{He\,{\sc ii}} $\lambda$ 4686 emission and Raman-scattered lines in LHA 115-S 18 could be explained by the presence of an accreting mass companion, as it was proposed by \citet{zic89}. However, this companion should be hot enough to produce highly ionized oxygen atoms. On the other hand, it is certainly interesting to remember that B[e] stars do not necessarily need to be really of spectral type B. Perhaps the central star of LHA 115-S 18 is hotter than the 25\,000 K reported in the literature and capable to emit O\,{\sc vi} photons. So, if these photons hit a dense disc that is neutral in hydrogen, we might expect to see Raman scattering. The possibility of a hotter central star in LHA 115-S 18 would place this object close to the LBV stars. An LBV-type status (with eruptions and ejections of large amounts of material) might account for the strong variability observed in the spectrum and the mass enhancement in the circumstellar environment. A mass ejection event in LHA 115-S 18 could be associated to the detection of P Cygni line profiles in a previous phase to the appearance of Raman-scattered lines. 

We want to stress that if these last scenarios are appropriate, LHA 115-S 18 would be the first B[e] supergiant or LBV star showing Raman-scattered lines. Therefore, in order to get further insight on the physical nature of LHA 115-S 18, a detailed description of the complete set of data, as well as its spectral and temporal variability, will be carried out in a forthcoming paper.

\section*{Acknowledgments}
We would like to thank the reviewer, H.M. Schmid, for his helpful comments which contributed to improve this Letter. M.K. acknowledges financial support from GA\,\v{C}R under grant number 209/11/1198. The Astronomical Institute Ond\v{r}ejov is supported by the project RVO:67985815. L.C. acknowledges financial support from the Agencia de Promoci\'on Cient\'ifica y Tecnol\'ogica (BID 1728 OC/AR PICT 0885), from CONICET (PIP 0300), and the Programa de Incentivos G11/089 of the Universidad Nacional de La Plata, Argentina. Financial support for International Cooperation of the Czech Republic
(M\v{S}MT, 7AMB12AR021) and Argentina (MINCyT-MEYS, ARC/11/10) is acknowledged. M.B.F. acknowledges Conselho Nacional de Desenvolvimento Cient\'ifico e Tecnol\'ogico (CNPq-Brazil) for the post-doctoral grant.

\bibliographystyle{mn2e}
\bibliography{biblio}

\begin{thebibliography}{}

\bibitem[\protect\citeauthoryear{{Allen}}{{Allen}}{1980}]{all80}
{Allen} D.~A.,  1980, MNRAS, 190, 75

\bibitem[\protect\citeauthoryear{{Azzopardi} \& {Breysacher}}{{Azzopardi} \&
  {Breysacher}}{1979}]{azz79}
{Azzopardi} M.,  {Breysacher} J.,  1979, A\&A, 75, 120

\bibitem[\protect\citeauthoryear{{Azzopardi}, {Breysacher} \&
  {Muratorio}}{{Azzopardi} et~al.}{1981}]{azz81}
{Azzopardi} M.,  {Breysacher} J.,    {Muratorio} G.,  1981, A\&A, 95, 191

\bibitem[\protect\citeauthoryear{{Azzopardi}, {Vigneau} \&
  {Macquet}}{{Azzopardi} et~al.}{1975}]{azz75}
{Azzopardi} M.,  {Vigneau} J.,    {Macquet} M.,  1975, A\&AS, 22, 285

\bibitem[\protect\citeauthoryear{{Belczy{\'n}ski}, {Miko{\l}ajewska}, {Munari},
  {Ivison} \& {Friedjung}}{{Belczy{\'n}ski} et~al.}{2000}]{bel00}
{Belczy{\'n}ski} K.,  {Miko{\l}ajewska} J.,  {Munari} U.,  {Ivison} R.~J.,
  {Friedjung} M.,  2000, A\&AS, 146, 407

\bibitem[\protect\citeauthoryear{{Birriel}}{{Birriel}}{2000}]{bir00a}
{Birriel} J.~J.,  2000, PhD thesis, University of Pittsburgh

\bibitem[\protect\citeauthoryear{{Birriel}, {Espey} \&
  {Schulte-Ladbeck}}{{Birriel} et~al.}{2000}]{bir00b}
{Birriel} J.~J.,  {Espey} B.~R.,    {Schulte-Ladbeck} R.~E.,  2000, ApJ, 545,
  1020

\bibitem[\protect\citeauthoryear{{Bonanos} \& {et al.,}}{{Bonanos} \& {et
  al.,}}{2010}]{bon10}
{Bonanos} A.~Z.,  {et al.,} 2010, AJ, 140, 416

\bibitem[\protect\citeauthoryear{{Henize}}{{Henize}}{1956}]{hen56}
{Henize} K.~G.,  1956, ApJS, 2, 315

\bibitem[\protect\citeauthoryear{{Kastner}, {Buchanan}, {Sahai}, {Forrest} \&
  {Sargent}}{{Kastner} et~al.}{2010}]{kas10}
{Kastner} J.~H.,  {Buchanan} C.,  {Sahai} R.,  {Forrest} W.~J.,    {Sargent}
  B.~A.,  2010, AJ, 139, 1993

\bibitem[\protect\citeauthoryear{{Kastner}, {Buchanan}, {Sargent} \&
  {Forrest}}{{Kastner} et~al.}{2006}]{kas06}
{Kastner} J.~H.,  {Buchanan} C.~L.,  {Sargent} B.,    {Forrest} W.~J.,  2006,
  ApJL, 638, L29

\bibitem[\protect\citeauthoryear{{Kenyon}}{{Kenyon}}{1986}]{ken86}
{Kenyon} S.~J.,  1986, {The symbiotic stars}

\bibitem[\protect\citeauthoryear{{Kraus} \& {Borges Fernandes}}{{Kraus} \&
  {Borges Fernandes}}{2005}]{kra05}
{Kraus} M.,  {Borges Fernandes} M.,  2005, in {Ignace} R.,  {Gayley} K.~G.,
  eds, The Nature and Evolution of Disks Around Hot Stars Vol.~337 of
  Astronomical Society of the Pacific Conference Series, {The outflowing disks
  of B[e] supergiants and unclassified B[e] stars}.
p.~254

\bibitem[\protect\citeauthoryear{{Kraus}, {Borges Fernandes} \& {de
  Ara{\'u}jo}}{{Kraus} et~al.}{2007}]{kra07}
{Kraus} M.,  {Borges Fernandes} M.,    {de Ara{\'u}jo} F.~X.,  2007, A\&A, 463,
  627

\bibitem[\protect\citeauthoryear{{Lamers}, {Zickgraf}, {de Winter}, {Houziaux}
  \& {Zorec}}{{Lamers} et~al.}{1998}]{lam98}
{Lamers} H.~J.~G.~L.~M.,  {Zickgraf} F.-J.,  {de Winter} D.,  {Houziaux} L.,
  {Zorec} J.,  1998, A\&A, 340, 117

\bibitem[\protect\citeauthoryear{{Leedj{\"a}rv}, {Burmeister},
  {Miko{\l}ajewski}, {Puss}, {Annuk} \& {Ga{\l}an}}{{Leedj{\"a}rv}
  et~al.}{2004}]{lee04}
{Leedj{\"a}rv} L.,  {Burmeister} M.,  {Miko{\l}ajewski} M.,  {Puss} A.,
  {Annuk} K.,    {Ga{\l}an} C.,  2004, A\&A, 415, 273

\bibitem[\protect\citeauthoryear{{Lindsay}}{{Lindsay}}{1955}]{lin55}
{Lindsay} E.~M.,  1955, MNRAS, 115, 248

\bibitem[\protect\citeauthoryear{{Massey} \& {Duffy}}{{Massey} \&
  {Duffy}}{2001}]{mas01}
{Massey} P.,  {Duffy} A.~S.,  2001, ApJ, 550, 713

\bibitem[\protect\citeauthoryear{{Morris}, {Eenens}, {Hanson}, {Conti} \&
  {Blum}}{{Morris} et~al.}{1996}]{mor96}
{Morris} P.~W.,  {Eenens} P.~R.~J.,  {Hanson} M.~M.,  {Conti} P.~S.,    {Blum}
  R.~D.,  1996, ApJ, 470, 597

\bibitem[\protect\citeauthoryear{{Nota}, {Pasquali}, {Drissen}, {Leitherer},
  {Robert}, {Moffat} \& {Schmutz}}{{Nota} et~al.}{1996}]{not96}
{Nota} A.,  {Pasquali} A.,  {Drissen} L.,  {Leitherer} C.,  {Robert} C.,
  {Moffat} A.~F.~J.,    {Schmutz} W.,  1996, ApJS, 102, 383

\bibitem[\protect\citeauthoryear{{Sanduleak}}{{Sanduleak}}{1978}]{san78}
{Sanduleak} N.,  1978, Information Bulletin on Variable Stars, 1389, 1

\bibitem[\protect\citeauthoryear{{Schmid}}{{Schmid}}{1989}]{sch89}
{Schmid} H.~M.,  1989, A\&A, 211, L31

\bibitem[\protect\citeauthoryear{{Shore} \& {Sanduleak}}{{Shore} \&
  {Sanduleak}}{1982}]{sho82}
{Shore} S.~N.,  {Sanduleak} N.,  1982, in {Y.~Kondo} ed., NASA Conference
  Publication Vol.~2338 of NASA Conference Publication, {The peculiar, luminous
  early-type emission line stars of the Magellanic clouds: A preliminary
  taxonomy}.
pp 602--605

\bibitem[\protect\citeauthoryear{{Shore}, {Sanduleak} \& {Allen}}{{Shore}
  et~al.}{1987}]{sho87}
{Shore} S.~N.,  {Sanduleak} N.,    {Allen} D.~A.,  1987, A\&A, 176, 59

\bibitem[\protect\citeauthoryear{{van Genderen} \& {Sterken}}{{van Genderen} \&
  {Sterken}}{2002}]{van02}
{van Genderen} A.~M.,  {Sterken} C.,  2002, A\&A, 386, 926

\bibitem[\protect\citeauthoryear{{Zickgraf}, {Wolf}, {Stahl} \&
  {Humphreys}}{{Zickgraf} et~al.}{1989}]{zic89}
{Zickgraf} F.-J.,  {Wolf} B.,  {Stahl} O.,    {Humphreys} R.~M.,  1989, A\&A,
  220, 206

\end{thebibliography}

\label{lastpage}
\end{document}